\documentclass[12pt,twoside]{article}
\usepackage{fleqn,espcrc1}

\title{Shell-Model Description of $\Lambda$ Hypernuclei}

\author{D. J. Millener\address{Brookhaven National Laboratory, Upton, 
    NY 11973, USA}}

\begin{document}

\maketitle

\bibliographystyle{unsrt}

\begin{abstract}
 Empirical data on the spectra of light hypernuclei, especially
the data from recent $\gamma$-ray experiments, is used
to constrain the parameters which govern the $p_Ns_\Lambda$
and $p_Np_\Lambda$ two-body matrix elements that enter into 
shell-model calculations.

\end{abstract}

\section{Introduction}

  Three hypernuclear $\gamma$-ray experiments were run in 1998.
The detector for KEK E419 and BNL E930 was the Hyperball.
The primary objective of KEK E419 was to measure the 
spacing of the the ground-state doublet in $^7_\Lambda$Li while that 
of BNL E930 was to measure the spacing of the excited-state doublet 
in $^9_\Lambda$Be built on the $2^+$ state of the $^8$Be core.
The objective of BNL E929, which used 72 NaI detectors and a live 
$^{13}$C target, was to measure the spacing of $p_{1/2}$ and $p_{3/2}$ 
$\Lambda$ states at $\sim 11$ MeV excitation energy in $^{13}_\Lambda$C. 

 While considerable progress has been made starting from a good
NN model (OBEP or OBEP + TBEP), applying symmetries, and fitting the
limited YN data set, the procedure does not lead to a reliable
YN interaction. Specifically, predictions for how the overall
binding energy is distributed between singlet and triplet, or
even and odd state, interactions vary widely. For example, six
NSC97 interactions have been constructed \cite{rij99} 
which give equally good fits to the YN data but which exhibit a wide
range of central spin-spin interaction strengths. On the other hand,
one can argue that many-body effective interaction theory provides
a sound connection between the free YN interaction and the effective
$\Lambda$N (and $\Lambda$NN) interaction for shell-model calculations.
Thus, precise hypernuclear data should strongly constrain models
of the free YN interaction.

  The $\Lambda$N$-\Sigma$N coupling is an important feature of 
YN interaction models.
Recently, Akaishi et al. \cite{akaishi} have calculated G-matrices
for a number of YN potential models for use in the model space of
$s$ orbits only. Then, the coupling between $s^3s_\Lambda$ and 
$s^3s_\Sigma$ configurations is simply $v=3/2\,^3g - 1/2\,^1g$
for $0^+$ and  $v=1/2\,^3g + 1/2\,^1g$ for $1^+$,  where $^3g$ and $^1g$  
are the relative s-state g-matrix elements for triplet and singlet states.
The $\Sigma$ admixture and energy shift for the lower state are 
$(v/\Delta E)^2$ and $v^2/\Delta E$, respectively ($\Delta E\sim 80$ MeV). 
The $^3g$ interaction dominates with the result that the energy shift
is substanial for the $0^+$ state and very small for the $1^+$ state.
The results for NSC97e and NSC97f bracket the experimental splitting
of $\sim 1.1$ MeV. For NSC97f, the $\Lambda$N and $\Lambda$NN contributions
are comparable ($v\sim 7.6$ MeV) for a total of 1.48 MeV.
At this meeting, E. Hiyama showed similar results from calculations
which include explicit $\Sigma$ degrees of freedom in a large model space.

 If one calculates the $\Lambda$N$-\Sigma$N
coupling for the $1/2^+$ ground-state of $^7_\Lambda$Li by asking
for the matrix element between $p^2(1^+;0)s_\Lambda$  and
$p^2(0^+;1)s_\Sigma$ in the LS limit, one gets 
$v=\sqrt{3}(^3g_{ps}- {^1g_{ps}})/2$. Because $g_{ps}\sim g_{ss}/2$, the
energy shift should be $\sim 1/12$ of that for the $0^+$
states in the A=4 hypernuclei. For the $3/2^+$ state, only non-central 
coupling interactions contribute and the shift should be very small indeed. 

 The last result justifies the initial neglect of effective
$\Lambda NN$ interactions for p-shell hypernuclei. Fetisov
\cite{fetisov99} has come to different conclusions using a
particular form of zero-range $\Lambda$NN interaction. 
 In the following sections, we consider how the new $\gamma$-ray
data affects the parametrizations of the $\Lambda$N effective
interaction put forward by MGDD \cite{mgdd85} and by 
FMZE \cite{fet91}. From this purely phenomenological point of view,
the spin-dependence for a $\Lambda$ in a $0s$ orbit interacting with
a p-shell core is specified by four radial integrals, conventionally
denoted by $\Delta$, $S_\Lambda$, $S_N$, and $T$ associated with
the operators $s_N.s_\Lambda$, $l_{N\Lambda}.s_\Lambda$,  
$l_{N\Lambda}.s_N$, and $3(\sigma_N.\widehat{r})
(\sigma_\Lambda.\widehat{r}) -\sigma_N.\sigma_\Lambda$. 

\begin{figure}[hb]
\setlength{\unitlength}{1in}
\thicklines
\begin{picture}(5,4.5)(-1,0)
\put(0.7,0.0){\line(1,0){2.5}}
\put(0.7,0.69){\line(1,0){2.5}}
\put(0.7,2.05){\line(1,0){2.5}}
\put(0.7,2.51){\line(1,0){2.5}}
\put(0.7,3.82){\line(1,0){2.5}}
\put(0.2,-0.02){0}
\put(0.2,0.67){0.71}
\put(0.2,2.03){2.05}
\put(2.8,1.7){{\Large $\tau = 5.8\pm 1.1$ ps}}
\put(-0.2,0.4){{\Large $E_x = 692(2)$ keV}}
\put(-0.2,1.5){{\Large $E_x = 2050(1)$ keV}}
\put(-0.2,3.0){{\Large $E_x = 3875(4)$ keV}}
\put(0.2,2.49){2.48}
\put(0.2,3.78){3.82}
\put(3.4,-0.02){$1/2^+$}
\put(3.4,0.67){$3/2^+$}
\put(3.4,2.03){$5/2^+$}
\put(3.4,2.49){$7/2^+$}
\put(3.4,3.78){$1/2^+;1$}
\put(4.2,-0.02){1.21}
\put(4.2,0.67){0.13}
\put(4.2,2.03){1.23}
\put(4.2,2.49){0.08}
\put(4.2,3.78){0.60}
\put(1.1,2.05){\circle*{0.1}}
\put(1.1,3.82){\line(0,-1){1.77}}
\put(0.9,2.1){96}
\put(1.5,2.05){\circle*{0.1}}
\put(1.5,3.82){\vector(0,-1){3.13}}
\put(1.3,2.1){4}
\put(1.1,3.82){\circle*{0.1}}
\put(0.9,3.70){52}
\put(1.5,3.82){\circle*{0.1}}
\put(1.3,3.70){48}
\put(1.9,2.51){\circle*{0.1}}
\put(1.9,2.51){\vector(0,-1){0.46}}
\put(1.8,2.59){83}
\put(2.3,2.51){\circle*{0.1}}
\put(2.3,2.51){\vector(0,-1){1.82}}
\put(2.2,2.59){17}
\put(2.7,0.69){\circle*{0.1}}
\put(2.55,0.77){100}
\put(2.2,-0.4){\Large $^{\bf 7}_{\bf \Lambda}${\bf Li}}
\put(4.0,-0.4){\Large {\boldmath $\sigma$}$(\pi^+,{\rm K}^+)$} 
\multiput(0.7,3.94)(0.3333,0.){8}{\line(1,0){0.1667}}
\put(2.7,4.00){$^5_\Lambda{\rm He}+d$}
\put(0.2,4.00){3.94(4)}
\put(1.1,2.05){\vector(0,-1){2.05}}
\put(2.7,0.69){\vector(0,-1){0.69}}
\end{picture}
\vspace*{10mm}
\caption{The bound-state spectrum of $^7_\Lambda$Li.}
\label{li7g}
\end{figure}
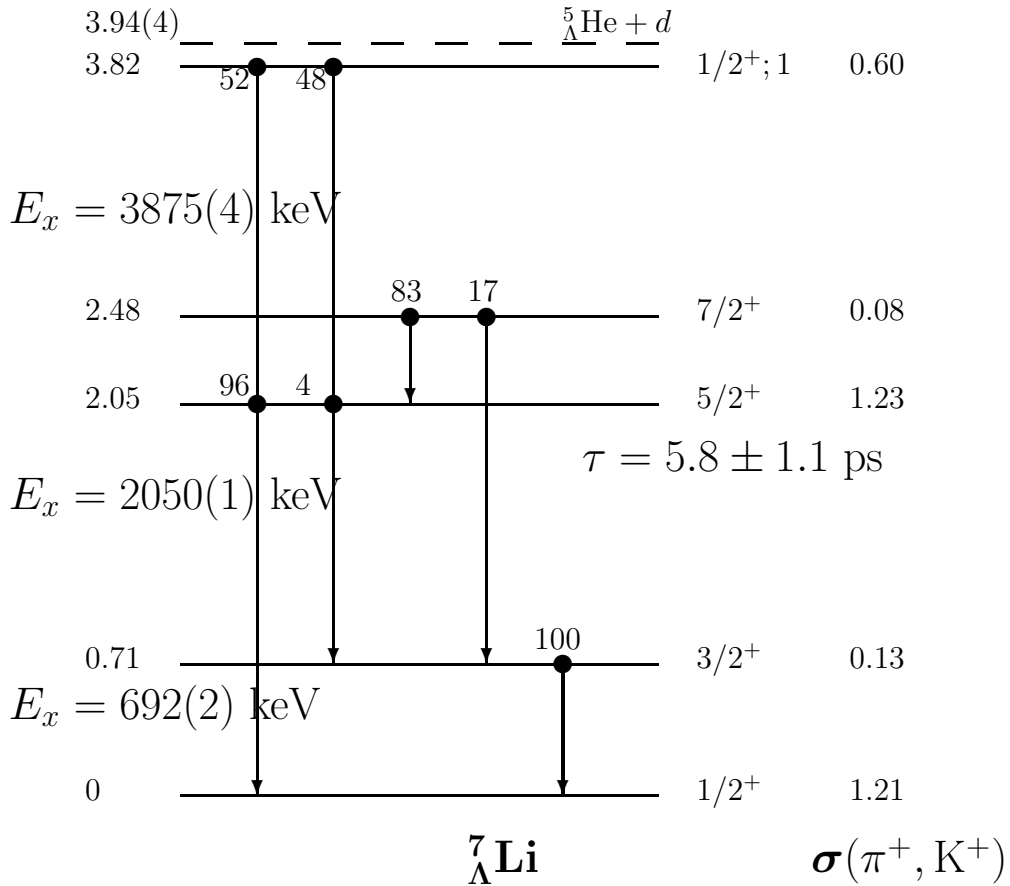

\section{$^{\bf 7}_{\bf\Lambda}$Li}

   The $\gamma$-ray energies \cite{tamura00} and the lifetime 
\cite{tanida00} from the $^7$Li$(\pi^+,{\rm K}^+\gamma)^7_\Lambda$Li 
reaction used in KEK E419 are shown in Fig.~\ref{li7g}.
The $\gamma$-ray  branching ratios  use the B(M1) and B(E2) values from 
the $^5_\Lambda{\rm He}+N+N$ model of Hiyama et al.~\cite{hiy99} and the
$(\pi^+,{\rm K}^+)$ cross sections integrated from $0^\circ -
15^\circ$ ($\mu$b) are also from Ref.~\cite{hiy99}.
The spectrum is from a shell-model calculation using a modified 
MGDD interaction with $S_N$ changed from $-0.08$ to $-0.47$ MeV 
to  bring the energy of the $5/2^+$ state in $^7_\Lambda$Li down to 
the measured value of 2.05 MeV, 
\begin{equation}
 \Delta =0.50\quad {\rm S}_\Lambda =-0.04\quad {\rm S}_{\rm N}
 =-0.47\quad T=0.04\ .
\label{eq:mgdd}
\end{equation}
At the same time, p-shell wave functions 
from a fit to 32 energy levels for $A=6-9$  are used 
for the $^6$Li core in place of the Cohen and Kurath wave functions 
used in MGDD \cite{mill00}. The main effect of a stronger one-body spin-orbit
component in the new p-shell interaction is to decrease the small $^3$S
admixture and increase the small $^1$P admixture in the dominantly
$^3$S ground state of $^6$Li \cite{mill00}. Similarly, the $^3$P
component in the $0^+;1$ state is substantially increased.

\begin{table}[b]
\caption{Breakdown of contributions to energy level spacings
in $^7_\Lambda$Li. For each spacing, the first line gives the 
coefficients (in MeV) for the contribution of each parameter
while the second gives the actual contributions (in keV). 
}
\begin{tabular*}{\textwidth}{@{}c@{\extracolsep{\fill}}ccccc}
\hline
 Spacing & $\Delta$ & S$_\Lambda$ & S$_{\rm  N}$ &  T & 
$\Delta E_{\Lambda N}$
 \\
\hline
 $3/2^+ - 1/2^+$ & 1.444 & 0.054 & ~~0.016 & $-0.271$  &  712 keV \\
 $\Delta E\sim\Delta E_{\Lambda N}$  & 722  & $-2$ &  $-8$ & $-11$  & \\
 & & & & & \\
  $5/2^+ - 1/2^+$  & 0.154  & $-1.105$  & 0.678 & 1.095 &  2050 keV \\
 $\Delta E \sim 2186 + \Delta E_{\Lambda N}$ & 77  & 44  & $-319$ & 44  &  \\
 & & & & & \\
 $1/2^+ - 1/2^+$  & 0.955  & 0.045 & 0.466 & $-0.057$ & 3819 keV  \\
 $\Delta E \sim 3563 + \Delta E_{\Lambda N}$ &  477  & $-2$  & $-219$ & 
$-2$  &  \\
 & & & & & \\
 $7/2^+ - 5/2^+$ & 1.311  & 2.141 & 0.024 &  $-2.324$ & 431 keV  \\
  $\Delta E\sim\Delta E_{\Lambda N}$ &  656  & $-86$  & $-11$ & $-93$  &  \\
\hline
\end{tabular*}
\label{tab:7lisep}
\end{table}

 The hypernuclear wave functions are very close to the weak-coupling limit.
The contribution of each component of the $\Lambda$N
interaction to the observed energy separations in $^7_\Lambda$Li
are given in Table \ref{tab:7lisep}. As expected, the energy splitting 
of the ground-state doublet is dominated by the spin-spin interaction.
A substantial negative value for $S_N$ is required to fit 
the excitation energy of the $5/2^+$ state.
The excitation energy of the $1/2^+$ T=1 state provides
another measure of $S_N$, the coefficient of which is controlled 
mainly by the $^3P$ component in the $0^+;1$ wave function. This
coefficient is very different from the original MGDD value of 0.081.

  In principle, the measurement of four energy separations,
the energies of two states based on excited states of the core and
two doublet separations, gives the four parameters controlling 
the spin dependence of the $p_Ns_\Lambda$ interaction. In practice,
$\Delta$, $S_N$, and one combination of $S_\Lambda$
and $T$, namely $T-S_\Lambda$, might be determined.

\section{\boldmath $s_\Lambda$ states in $^{\bf 9}_{\bf\Lambda}$Be, 
$^{\bf 12}_{\bf\ \Lambda}$C, and $^{\bf 13}_{\bf\ \Lambda}$C}

 There is new data on each of these hypernuclei from experiments at KEK 
and BNL and it is of interest to see whether there is consistency
with the preceding analysis of the KEK E419 data on $^7_\Lambda$Li.
In this analysis, the $p_Ns_\Lambda$ matrix elements were treated as 
parameters. However, the radial integrals depend on the size of the 
hypernucleus and to study this effect the
YNG interactions of Yamamoto et al. \cite{yamamoto94}, in which
nuclear matter G-matrix elements for various free YN interaction models
are simulated by a number of Gaussians with different ranges ($r^2$
times Gaussian for tensor interactions), are used.  In particular,
the parametrization \cite{yamamoto} of the NSC97f interaction 
\cite{rij99} for $k_F=1$ fm$^{-1}$ is used with Woods-Saxon radial
wave functions. In the first instance, the strengths of each component 
are scaled to reproduce the parameters of Eq. (\ref{eq:mgdd}).
To do this the singlet and triplet central interaction strengths
have to be slightly reduced and increased, respectively, to reduce
$\Delta$ from the NSC97f value. The ALS component has to be increased
by a factor of $\sim 3$ and the tensor component has to be reduced
somewhat in strength. Results are shown in Table~\ref{tab:paramws},
where it can be seen that matrix elements are larger for more
tightly bound neutron wave functions.
 
 The value $\Delta = 0.48$ MeV chosen for $^7_\Lambda$Li gives
improved agreement with the measure ground-state doublet splitting
of 692 keV. The smaller magnitudes for $S_\Lambda$ and $T$ give
better agreement with the new data from BNL E930 on the
doublet splitting in $^9_\Lambda$Be, to be discussed next. As
can be seen from Table~\ref{tab:7lisep}, these reductions
require a small reduction in the magnitude of $S_N$ to maintain
the excitation energy predicted for the $5/2^+$, which will then
improve the predicted energy for the $1/2^+;1$ state. The
spacing of the $7/2^+ - 5/2^+$ doublet would also increase,
perhaps too close to 511 keV to be directly measureable.

\begin{table}[b]
\caption{$p_ns_\Lambda$ parameters as a function of $A$ for
Woods-Saxon wave functions. For nucleons $r_0 = 1.25$ fm, $a=0.6$
fm; for $\Lambda$'s $r_0 = 1.128 + 0.439A^{-2/3}$ fm, $a=0.6$ fm.
Energies are in MeV and lengths are in fm.  
}
\begin{tabular*}{\textwidth}{@{}l@{\extracolsep{\fill}}rrrrr}
\hline
  & $^7_\Lambda$Li & $^9_\Lambda$Be  & $^{12}_\Lambda$C &  
$^{13}_\Lambda$C &$^{16}_\Lambda$O   \\
\hline
 B$_\Lambda(0s)$ & 5.58 & 6.73 & 10.80 & 11.67 & 13.0 \\
 B$_n(0p)$ & 5.56 & 18.90 & 13.12 & 18.72 & 16.31 \\
 $\langle r^2\rangle^{1/2}(0s_\Lambda)$ & 2.63 & 2.55 & 2.33 & 2.32 & 2.32 \\
 $\langle r^2\rangle^{1/2}(0p_N)$       & 2.86 & 2.34 & 2.62 & 2.50 & 2.66 \\
 $\Delta$    & 0.480    & 0.619    & 0.550    & 0.591    & 0.521 \\
  $S_N$      & $-0.430$ & $-0.549$ & $-0.508$ & $-0.545$ & $-0.461$ \\
 $S_\Lambda$ & $-0.010$ & $-0.013$ & $-0.012$ & $-0.013$ & $-0.011$ \\
 $T$         & $0.021$  & $0.029$  & $0.025$  & $0.027$  & $0.023$ \\
\hline
\end{tabular*}
\label{tab:paramws}
\end{table}

\begin{table}[t]
\caption{$^{9}_{\Lambda}$Be:\  $3/2^+ - 5/2^+$ separation} 
\begin{tabular*}{\textwidth}{@{}c@{\extracolsep{\fill}}cccc}
\hline
  $\Delta$ & S$_\Lambda$ & S$_{\rm  N}$ &  T & $\Delta E$  \\
\hline
  $-0.036$ & $-2.463$ & $0.002$ & $0.985$ & 40 keV \\
   $-22$  &  32  &  $-1$  &   29  &     \\
\hline
\end{tabular*}
\label{tab:9besep}
\end{table}

   The result from BNL E930 reported at this meeting by Tamura and
by Akikawa is that the splitting of the $3/2^+$, $5/2^+$ doublet in 
$^9_\Lambda$Be is only $\sim 32$ keV, updating the limit of 
$< 100$ keV from the work of May et al. \cite{may83} with NaI detectors. 
As can be seen from Table~\ref{tab:9besep}, the small $S=1$ amplitudes 
($\sim 4$\% intensity) in the $^8$Be $2^+$ wave function (necessary to 
account for $^8$Li and $^8$B $\beta$ decay) lead to a substantial 
contribution from the $\Lambda$N tensor interaction. For $S_\Lambda
< 0$ and $T>0$, the measurement implies that the magnitudes of these
parameters are restricted to be very small.

  There is strong evidence from both $^{12}_{\ \Lambda}$C and
$^{13}_{\ \Lambda}$C that supports the substantial negative
value for $S_N$ deduced from $^7_\Lambda$Li.

 At this meeting, Sakaguchi reported an energy of 4.915(33) MeV from 
BNL E929 for the $\gamma$-ray from the first $3/2^+$ state of 
$^{13}_{\ \Lambda}$C. This is in agreement the value 4.89(7) MeV 
deduced from a high-statistics $^{13}$C$(\pi^+,K^+)^{13}_{\ \Lambda}$C 
spectrum from KEK E336 \cite{hashimoto}, in which the ground-state 
and $3/2^+$ peaks are well resolved. The deviation from the 4.439 MeV 
energy of the $2^+$ core can be attributed mainly to $S_N$, as can be 
seen from Table \ref{tab:13cs} (parameters from Table~\ref{tab:paramws}). 
For comparison, the MGD excitation energy was 4.49 MeV.
 
\begin{table}[h]
\caption{$^{13}_{\ \Lambda}$C:\  $3/2^+ - 1/2^+$ separation 
$\Delta E = 4439 + \Delta E_{\Lambda N}$}
\begin{tabular*}{\textwidth}{@{}c@{\extracolsep{\fill}}cccc}
\hline
   $\Delta$ & S$_\Lambda$ & S$_{\rm  N}$ &  T & $\Delta E$  \\
\hline
  $-0.050$  & $-1.450$ & $-0.861$ & $-1.104$ &  4823 keV \\
    $-30$  & 19  & 469 & $-30$  &   \\
\hline
\end{tabular*}
\label{tab:13cs}
\end{table}

 The energies of the excited $1^-$ states in $^{12}_\Lambda$C
are also raised from the unperturbed core energies. A preliminary
analysis of $(\pi^+,K^+)$ data from KEK E336 \cite{hashimoto}
gives 2.71 MeV and 6.05 MeV for the $1^-_2$ and $1^-_3$ levels.
The corresponding numbers from KEK E369 \cite{nagae}, which
achieved an energy resolution of 1.45 MeV, are 2.54 MeV and 6.17
MeV. As can be seen from Table~\ref{tab:12csep},
the largest contributions are again from $S_N$ with substantial
contributions from $\Delta$.
 
\begin{table}[ht]
\caption{Energy separations in $^{12}_{\ \Lambda}$C}  
\begin{tabular*}{\textwidth}{@{}c@{\extracolsep{\fill}}ccccc}
\hline
 Spacing & $\Delta$ & S$_\Lambda$ & S$_{\rm  N}$ &  T & $\Delta E$  \\
\hline
 $1^-_2 - 1^-_1$  & 0.331  & 1.147 & $-0.913$ & 0.670 &  2610 keV \\
 $\Delta E\sim 2.000 +\Delta E_{\Lambda N}$ &  182  & $-14$  & 464 & 17  & \\
 & & & & & \\
 $1^-_3 - 1^-_1$  & 0.376  & $-0.388$ & $-1.339$ & 0.464 &  5685 keV \\
 $\Delta E\sim 4.804 + \Delta E_{\Lambda N}$ &  207  & 5  & 680 & 12  &  \\
 & & & & & \\
 $2^-_1 - 1^-_1$  & 0.474 & 1.510 & ~~0.031 & $-2.092$ & 167 keV  \\
  $\Delta E\sim \Delta E_{\Lambda N}$ & 261 &  $-18$  &  $-16$ & $-52$  &  \\
\hline
\end{tabular*}
\label{tab:12csep}
\end{table}

  A number of $\gamma$ rays in $^{12}_{\ \Lambda}$C could be
measured in future runs of BNL E930 at either 0.9 GeV/c or 1.8 GeV/c
depending in part on the primary beam intensity available to the D6
beamline. These include the spacing of the ground-state doublet, 
either directly if the spacing is large enough to preclude 
predominantly weak decay of the upper level or indirectly through 
transitions from the $1^-_2$ level. The simplest weak-coupling estimate
gives 280 ps for the partial electromagnetic lifetime of the $2^-_1$
level for the separation in Table \ref{tab:12csep}. The parameters 
of Eq. (\ref{eq:mgdd}) give a spacing of only 71 keV so that weak decay
would dominate.

\section{\boldmath $p_\Lambda$ states in $^{\bf 13}_{\ \bf\Lambda}$C}

 The $p_np_\Lambda$ central interaction can be parametrized  for
$S=0$ and $S=1$ separately as~\cite{auer83} 
\begin{equation}
 V_{N\Lambda} = (F^{(0)}+F^{(2)}Q_N.Q_\Lambda)(1-\varepsilon
+\varepsilon P_x) \label{eq:qqint}
\end{equation}
where $F^{(2)}/F^{(0)}$ characterizes the range of the interaction and
$\varepsilon$ characterizes the space-exchange component.
The NSC97 interactions exhibit strong short-range repulsion in
relative p waves, corresponding to $\varepsilon\sim 1$ in the 
parametrization of Eq. (\ref{eq:qqint}). Interactions with
different space-exchange properties can be made from the YNG
interactions by varying the strengths of the even and odd state
interactions separately while maintaining the same overall
attraction. 

 The essential structure of the lowest $p_\Lambda$ states in 
$^{13}_{\ \Lambda}$C, built on the lowest $0^+$ and $2^+$ states of the 
$^{12}$C core, is illustrated in Fig.~12 of Ref.~ \cite{auer83}. 
The $2^+\otimes p_\Lambda$ states appear in the order ${\cal L}$ =
2,3,1 where ${\cal L} = J_{core} + l_\Lambda$ under 
the action of the $Q.Q$ interaction and doublets form as a
the result of coupling the $\Lambda$ intrinsic spin. A 
space-exchange interaction pushes the $0^+\otimes p_\Lambda$
and $2^+\otimes p_\Lambda$ states with ${\cal L}$ = 1 apart.

The two $1/2^-$ states with ${\cal L}$ = 1 are seen strongly via $\Delta L=0$
transitions in the $(K^-,\pi^-)$ reaction near $0^\circ$, the upper one
most strongly because it tends towards the [441] spatial symmetry
of the $^{13}$C target. The separation is observed to be 6.0(4) MeV
\cite{may81}. The NSC97f interaction, which has a very strong space-exhange
component, gives a spacing of 10 MeV when harmonic oscillator wave 
functions are used to reproduce the parameters of Eq.~(\ref{eq:mgdd}).
From BNL E929, the lowest $1/2^-$,$3/2^-$ doublet is at $E_x\sim 11$ MeV
so the the $0p_\Lambda$ separation energy is only 0.67 MeV. The
Woods-Saxon parameters used in Table~\ref{tab:paramws} give an
rms radius for the $0p_\Lambda$ orbit of 4.53 fm and the 
mismatch with the deeply bound nucleon orbits leads to a very
substantial reduction in the $p_Np_\Lambda$ matrix elements. 
The separation of the $1/2^-$ states drops to 7.7 MeV, which is
still too large. The interaction used for Table~\ref{tab:paramws}
has an attractive odd-state interaction of about half the strength of the
even-state interaction ($\varepsilon\sim 0.25$) and reproduces the
experimentally measured $1/2^-$ separation of 6.0 MeV. The ratio
of cross sections for the $1/2^-$ states increases as the strength of 
the space-exchange interaction increases. Experiment ($\sim 5.5$) 
and theory are in good agreement for the interaction which reproduces the
energy separation.

 The $3/2^-_1$ and $5/2^-_2$ states are seen strongly
in the $(\pi^+,K^+)$ reaction or the $(K^-,\pi^-)$ reaction at larger
angles and the $1/2^-_2-5/2^-_2$ separation has been measured to be
1.7(4) MeV \cite{may81}. The NSC97f and  $\varepsilon\sim 0.25$ interactions
give 2.9 and 1.3 MeV, respectively, again favoring the interaction with
some p-state attraction. Such a feature is present in the ESC99 
interaction described by Rijken at this meeting.

 BNL E929 has given a small separation for the lowest $1/2^-$ and 
$3/2^-$ states of $^{13}_{\ \Lambda}$C at 11 MeV
of 152(54) keV, with an estimated systematic error of 36 keV. 
To a first approximation, these states may be regarded
as $p_{1/2}$ and $p_{3/2}$ $\Lambda$ single-particle states with
the $^{12}$C ground state as a core and their separation was thought to
give a measurement of the $p_\Lambda$ spin-orbit splitting. 
However, the spin-spin, spin-orbit, and tensor interactions all make 
substantial contributions to the splitting, as can be
seen from Table \ref{tab:13csep} (the splitting has yet to be
decomposed into the individual contributions for the Woods-Saxon case).
Note that models which use $\alpha$-particle cores \cite{hiyama00}
allow only the spin-orbit interactions to contribute to the
splittings in $^{13}_{\ \Lambda}$C and $^9_{\Lambda}$Be.

\begin{table}[h]
\caption{$^{12}$C($0^+)\otimes p_\Lambda$\  $1/2^- - 3/2^-$\  separation
in $^{13}_\Lambda$C for harmonic oscillator wave functions.}
\begin{tabular*}{\textwidth}{@{}c@{\extracolsep{\fill}}ccc}
\hline
spin-spin & spin-orbit & tensor &  total   \\
\hline
 $+42$ keV & $+280$ keV & $-215$ keV & 107 keV   \\
\hline
\end{tabular*}
\label{tab:13csep}
\end{table}

 The spin-spin and tensor contributions arise from $S=1$ components in
$^{12}$C wave functions. For the CKPOT wave function used for the core,
there is a substantial (16\%) L=1,S=1 component with [431] symmetry
mixed into the dominant (79\%) [44] L=0,S=0 component for the $0^+$
state. The spin-orbit contribution to the separation of the diagonal 
$0^+\otimes p_\Lambda$ matrix elements gets  240 keV from $s^4p_\Lambda$ 
and 80 keV from $p^8p_\Lambda$. The $2^+\otimes p_\Lambda$ admixtures
are at the level of $\sim 5$\% and act in a somewhat different
manner for $1/2^-$ and $3/2^-$ \cite{auer83}. The  tensor contribution
comes $\sim 2/3$ from even state and $\sim 1/3$ odd state interactions,
a result which is specific to the NSC97f interaction but typical of
a number of the available YN interactions.
The fact that there is substantial cancellation
between the spin-orbit and tensor contributions is in contrast to
$^9_\Lambda$Be where spin-orbit and tensor contributions
add rather than subtract in the $3/2^+$, $5/2^+$ separation (Table 
\ref{tab:9besep}). The interaction used in  Table~\ref{tab:paramws},
and which gives a good description of other features of the
$p_\Lambda$ states, gives a splitting of 33 keV for Woods-Saxon
wave functions with a much smaller contribution from the
spin-orbit interaction than is shown in Table \ref{tab:13csep}.

\section{Remarks}

 The new results on $s_\Lambda$ states of $^7_\Lambda$Li,
$^9_\Lambda$Be,  $^{12}_{\ \Lambda}$C, and $^{13}_{\ \Lambda}$C
are consistent with substantial magnidues for $\Delta$ and $S_N$
and small values for $S_\Lambda$ and $T$.

 The $p_Np_\Lambda$ matrix elements are sensitive to more
features of the underlying $\Lambda$N interaction than are the
$p_Ns_\Lambda$ matrix elements. Quantities such as the space-exchange
character, the Q.Q component, and the even-state tensor
interaction all play a role for $p_\Lambda$ states
in $^{13}_{\ \Lambda}$C. In particular, the magnitude of the 
space-exchange component of the effective $\Lambda$N interaction is
restricted to be quite small. This is in agreement with an analysis 
of $\Lambda$ single-particle
energies by Usmani and Bodmer \cite{usmani99}, who also find $\varepsilon\sim
0.25$.

 The object of future $(K^-,\pi^-\gamma)$ runs of BNL E930
is to obtain a much more complete the set of information on the 
$p_Ns_\Lambda$ interaction. An important target is $^{16}$O because 
the ground-state doublet splitting in $^{16}_{\ \Lambda}$O is very 
sensitive to the tensor matrix element $T$ \cite{mgdd85}. It will
probably have to be measured as the difference of energies of the $\sim 6$ MeV 
$\gamma$-rays deexciting the $1^-_2$ level. It may also be possible
to observe $\gamma$-ray transitions in $^{15}_{\ \Lambda}$N
following proton emission from higher-energy states 
in $^{16}_{\ \Lambda}$O. As mentioned previously, studies with 
$^{12}$C and $^7$Li targets are also under consideration. Measurements
for a range of nuclei throughout the $p$ shell should both overdetermine
the set of spin-dependent parameters and make it possible to
test for variations with nuclear size and the empirical need for
a $\Lambda$NN interaction. 

\section*{Acknowledgement}

 This work has been supported by the U.S. Department of Energy
under Contract No. DE-AC02-98CH10886.


\begin{thebibliography}{10}

\bibitem{rij99} Th.A. Rijken, V.G.J. Stoks, and Y. Yamamoto,
Phys. Rev. C {\bf 59} (1999) 21.

\bibitem{akaishi} Y. Akaishi {\it et al.}, Phys. Rev. Lett. {\bf 84},
 (2000) 3539.

\bibitem{fetisov99} V.N. Fetisov, JETP Lett. {\bf 70} (1999) 233.

\bibitem{mgdd85} D.J. Millener, A. Gal, C.B. Dover, and R.H. Dalitz,
Phys. Rev. C {\bf 31} (1985) 499.

\bibitem{fet91} V.N. Fetisov, L. Majling, J. \v{Z}ofka, and R.A. Eramzhyan, 
  Z. Phys. A {\bf 339} (1991) 399.

\bibitem{tamura00} H. Tamura {\it et al.}, Phys. Rev. Lett. {\bf 84},
 (2000) 5963.

\bibitem{tanida00} K. Tanida {\it et al.}, Phys. Rev. Lett., in press.

\bibitem{hiy99} E. Hiyama, M. Kamimura, K. Miyazaki, and T. Motoba, 
Phys. Rev. C {\bf 59} (1999) 2351.

\bibitem{mill00} D.J. Millener, in Strangeness Nuclear Physics, eds. Il-T.
Cheon, S.W. Hong, and T. Motoba (World Scientific, 2000) p. 81.

\bibitem{yamamoto94} Y. Yamamoto {\it et al.}, Prog. Theor. Phys. Suppl.
{\bf 117} (1994) 361.

\bibitem{yamamoto} Y. Yamamoto, private communication.

\bibitem{may83} M. May {\it et al.}, Phys. Rev. Lett. {\bf 51} (1983) 2085.
 
\bibitem{hashimoto} O. Hashimoto,  Nucl. Phys. A {\bf 639} (1998) 93c.

\bibitem{nagae} T. Nagae, in Strangeness Nuclear Physics, eds. Il-T.
Cheon, S.W. Hong, and T. Motoba (World Scientific, 2000) p. 110.

\bibitem{auer83}  E.H. Auerbach {\it et al.}, Ann. Phys. (NY) {\bf 148} 
 (1983) 381.

\bibitem{may81} M. May {\it et al.}, Phys. Rev. Lett. {\bf 47} (1981) 1106;
 E.H. Auerbach {\it et al., ibid.} {\bf 47} (1981) 1110.

\bibitem{hiyama00} E. Hiyama {\it et al.},
Phys. Rev. Lett. {\bf 85} (2000) 270.

\bibitem{usmani99} Q.N. Usmani and A.R. Bodmer, Phys. Rev. C {\bf 60},
 055215 (1999).

\end{thebibliography}
\end{document}